\documentclass[article,twocolumn,
 amsmath,amssymb
 prl,
]{revtex4-1}

\usepackage{graphicx}
\usepackage{mathtools}
\usepackage{dcolumn}
\usepackage{bm}
\usepackage{amsthm}
\usepackage{amsfonts}
\usepackage{hyperref}
\usepackage[dvipsnames]{xcolor}
\usepackage{subfig}
\usepackage{float}
\begin{document}

\title{The Quantum Skin Hall Effect}

\author{Yuhao Ma}
\author{Taylor L. Hughes}
\affiliation{ Department of Physics and Institute for Condensed Matter Theory$,$\\
 University of Illinois at Urbana-Champaign$,$ Illinois 61801$,$ USA}

\begin{abstract}
The skin effect, which is unique to non-Hermitian systems, can generate an extensive number of eigenstates localized near the boundary in an open geometry. Here we propose that in 2D and 3D other quantities besides charge density are susceptible to the skin effect. We show that 2D and 3D models that are a hybrid between topological insulators and skin-effect systems can have a topological skin effect where an extensive number of topological modes, and the corresponding bulk topological invariant, are pinned to the surface. A key example, which we call the \emph{quantum skin Hall effect} is constructed from layers of Chern insulators and exhibits an extensive Hall conductance and number of chiral modes bound to surfaces normal to the stack of layers. The same procedure is further extended to other symmetry classes to illustrate that a variety of 1D and 2D topological invariants ($\mathbb{Z}$ or $\mathbb{Z}_2)$ are subject to the skin effect. Indeed, we also propose a hybrid 2D system that exhibits an extensive number of topological corner modes and may be more easily realized in meta-material experiments.
\end{abstract}

\maketitle

The discovery of the skin effect\cite{yao2018edge,zhang2019correspondence, okuma2020topological,lee2019hybrid,lee2019anatomy,ghatak2019observation,helbig2019observation,brandenbourger2019non,zhou2020non,xiao2020non,hofmann2020reciprocal,song2019non} is a striking confirmation of the unusual character of non-Hermitian (NH) systems\cite{bender1998real,bender1999pt,bender2007making}. 
The conventional NH skin effect manifests in a macroscopic number of modes localized on the boundary of a system. In this article we propose a 3D \emph{quantum skin Hall effect} where, instead of an extensive number of localized skin modes, we observe an extensive quantized Hall conductance pinned to the surface. Our effect can be generalized to show that a 3D system can exhibit a type of topological skin effect where an extensive amount of a variety of 2D topological invariants, and their associated edge states, can be localized on the boundary. Since the NH skin effect has been successfully observed in  mechanical\cite{ghatak2019observation,brandenbourger2019non,zhou2020non}, optical\cite{xiao2020non}, and electronic systems\cite{helbig2019observation,hofmann2020reciprocal}, we expect our results may be experimentally realized imminently.

Recently there have been a number of investigations into the topological properties of NH systems, both theoretically\cite{hu2011,song2019non,gong2018topological,kawabata2019topological,hamazaki2019non,lee2019hybrid,zhu2014pt,yin2018geometrical,lieu2018topological,kawabata2019classification,kawabata2019symmetry,zhou2019periodic,budich2019symmetry,lee2019anatomy,deng2019non,yokomizo2019non,song2019non,song2019non,lee2016anomalous,shen2018topological,hu2011absence,kawabata2018parity,okuma2020topological,zhang2019correspondence,yao2018edge,yao2018non} and experimentally\cite{xiao2017observation,weimann2017topologically,zhao2018topological,parto2018complex,zhan2017detecting,xiao2020non,helbig2019observation,hofmann2020reciprocal,xiao2020non,zhou2020non,brandenbourger2019non,ghatak2019observation}. Indeed, non-Hermiticity can be realized in various physical contexts: open systems\cite{konotop2016nonlinear,el2018non}, optics\cite{feng2017non,miri2019exceptional,malzard2015topologically,mochizuki2016explicit}, quantum critical phenomena\cite{kawabata2017information,nakagawa2018non,xiao2019observation}, and disordered or correlated systems\cite{hamazaki2019non,yoshida2019symmetry,zyuzin2018flat}.
Despite intense effort, integrating non-Hermiticity and topology has proved challenging, and many open questions remain. The key difficulty arises because of the dichotomy in the bulk-boundary correspondence. On one hand, the celebrated deterministic connection between bulk topological invariants and robust boundary modes is the cornerstone of topological insulator phenomena\cite{hasan2010colloquium,bernevig2013topological}. On the other hand, NH systems have an extreme sensitivity to boundary conditions such that most bulk properties in periodic boundary conditions (PBC) have no bearing on the phenomenology of systems with a boundary (OBC). The NH skin effect is the paradigm example of the boundary-condition sensitivity.

Here we take a hybrid approach to combining non-Hermiticity and topology. We consider 2D/3D systems built from layers of 1D/2D topological insulators that have non-Hermitian coupling between layers. In the Hermitian limit these systems will exhibit a conventional bulk-boundary correspondence where the boundaries of the layers will exhibit topological modes that can hybridize and spread across the side surfaces\cite{balents1996chiral}. Once non-Hermiticity is added, these topological boundary modes will experience a NH skin effect in the transverse direction that will pin all of the modes on the top and/or bottom surfaces. As we will see, when the 2D layer is a Chern insulator this effect will generate an extensive number of chiral edge modes trapped at the top or bottom of a sample, and an associated extensive, quantized Hall conductance. This construction can be generalized to other symmetry classes as well, e.g., time-reversal invariant(TRI) systems built from layers having a non-trivial $\mathbb{Z}_2$ invariant.

\begin{figure}
    \centering
    \includegraphics[width=\columnwidth]{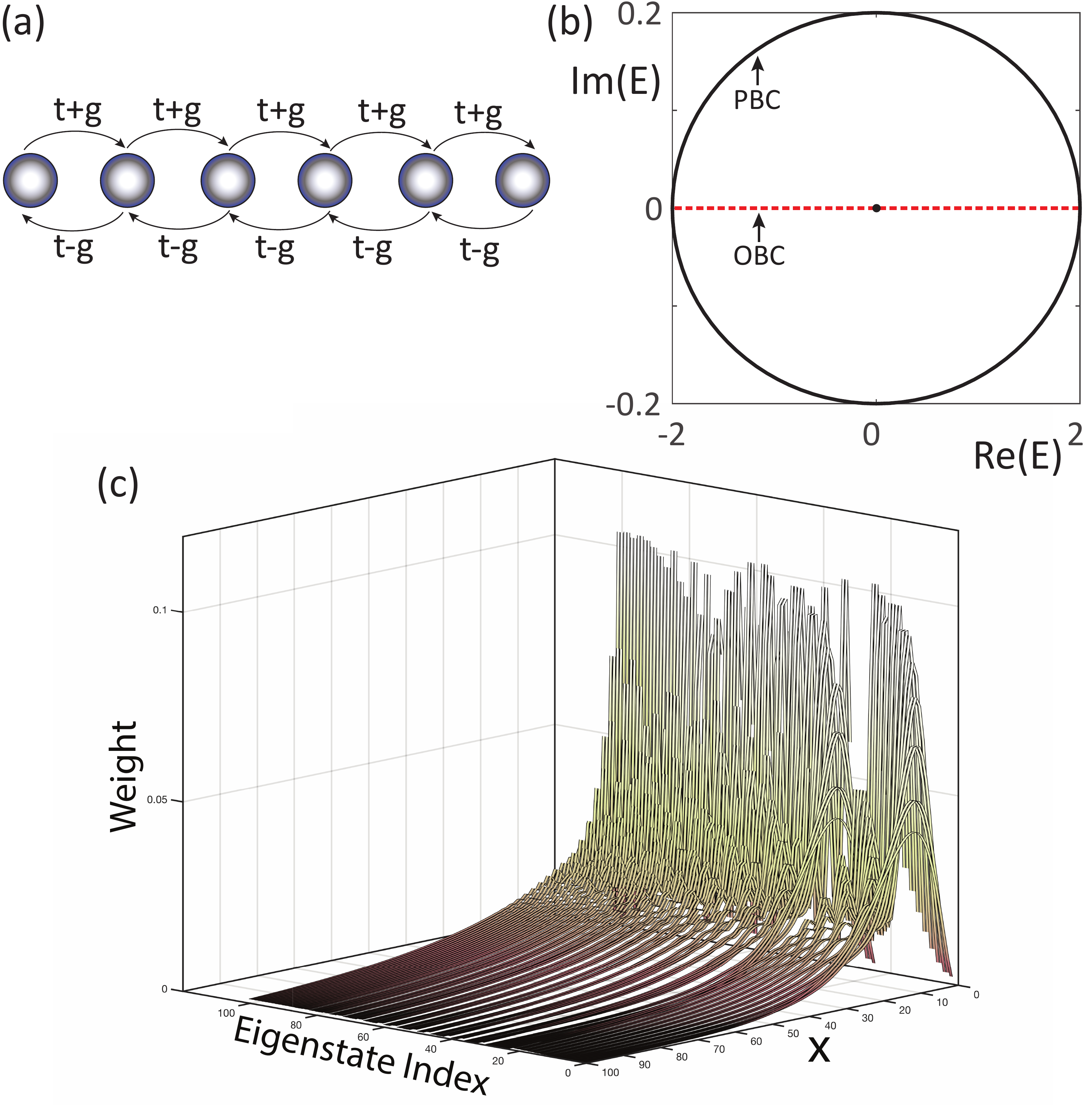}
    \caption{(a) The Hatano-Nelson tight-binding model. (b) PBC bulk spectrum (black loop) and OBC  spectrum(red dotted line) on the complex-energy plane, illustrating the high sensitivity of the system to its boundary conditions. (c) Spatial profile of all eigenstates vs lattice site index $x$ in the OBC spectrum, showing the NH skin effect. Parameters: $t=1$, $g=-0.1$.}
    \label{fig:HN}
\end{figure}

We begin by reviewing the translationally invariant 1D Hatano-Nelson (HN) model\cite{hatano1996localization, hatano1997vortex} (illustrated in Fig. \ref{fig:HN}(a)). The HN model is a 1D single-band tight-binding model where non-Hermiticity is introduced via unequal hopping amplitudes in the left and right directions: \begin{equation}
 H_{HN}:=\sum_{i} t_{+}\hat{c}^{\dagger}_{i+1}\hat{c}_{i}+t_{-}\hat{c}^{\dagger}_{i}\hat{c}_{i+1},
\end{equation}\noindent where $t_{\pm}=t\pm g$, $t > g$, and $ g\in \mathbb{R}$. For PBC the Bloch Hamiltonian is $H_{HN}(k)=t_{+}e^{ik}+t_{-}e^{-ik},$ which, for any $g\ne 0,$ has a complex spectrum in the shape of an ellipse that encloses the origin of the complex energy plane (shown in Fig. \ref{fig:HN}(b)). 

A key development of the NH bulk-boundary correspondence is that the winding number $W(E)$ of the PBC spectrum around a complex energy $E$ characterizes a so-called `point-gap topology.' If $W(E)$ is non-vanishing it signals the existence of the NH skin effect in OBC\cite{okuma2020topological, zhang2019correspondence}. The winding number is unique to NH systems, as the spectrum of any Hermitian system lies completely on the real axis, making $W(\textit{E})$ ill-defined. Specifically for the HN model, for any $\textit{E} \in \mathbb{C}$ inside PBC spectral ellipse, the winding number is $W(\textit{E}) = {\rm{sgn}}(g)$.  We see that for $W(E)$ to change its value $g$ has to change its sign, during which the system will pass through the Hermitian limit such that the point gap closes. The sign of $W(E)$ indicates the side on which the skin modes are localized. Heuristically,  if the sign of $g$ is chosen to be negative, the hopping from right to left dominates and states flow to the left, and vice-versa. As depicted in Fig. \ref{fig:HN}(c), for $g<0$ every state in the OBC spectrum has its weight exponentially localized on the left half of the chain with a decay length controlled by the parameter $t_{-}/t_{+}$.

Having seen that the NH skin effect generates an extensive charge/mode density on the surface, it begs the question about what other physical quantities might be susceptible to a skin effect. Our approach to answering this question can be illustrated with a simple example. Suppose we have a stack of decoupled 2D topological insulator layers, all having a non-trivial topological invariant and associated edge states. If we couple them in the stacking direction then, in the order of increasing inter-layer (Hermitian) coupling, we might arrive at a weak topological insulator(WTI)\cite{balents1996chiral,fu2007topological}, a topological semi-metal\cite{burkov2011topological,ramamurthy2015patterns}, or a trivial insulator. However, if we modify the inter-layer coupling using the HN model as a guide, we might expect that the extensive number of topological edge states, and the associated bulk topological phenomena, might be pushed to the top or bottom of the stack in a new type of NH skin effect. 

\begin{figure}[t!]   
\centering
\includegraphics[width=\columnwidth]{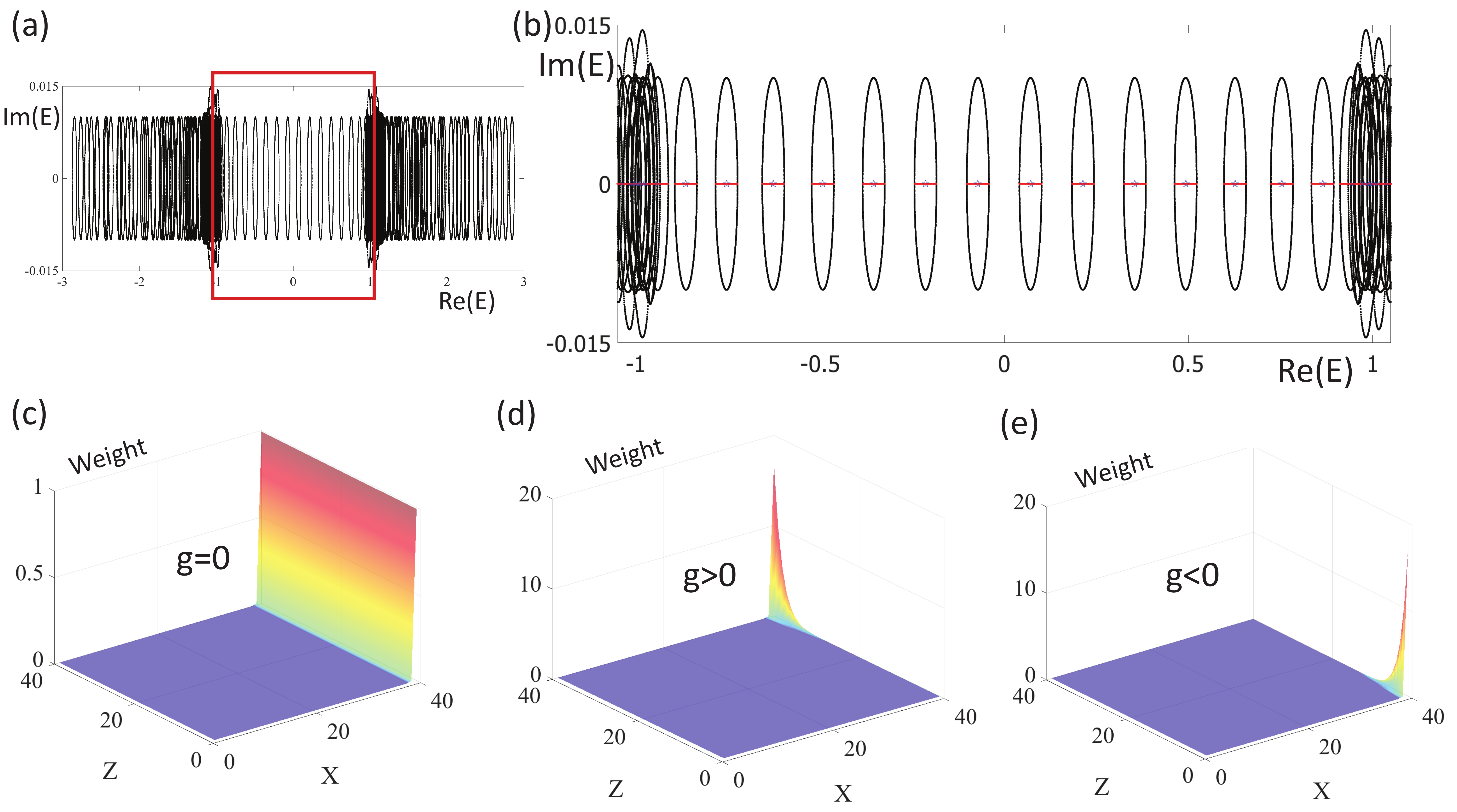}
  \caption{(a) Complex energy spectrum with PBC in $z$, and OBC in $x,y$. (b) Zoom in on the boundary state region from (a) in red box. Blue stars have $t=g=0,$ red lines have $t\neq 0, g=0$, and black lines have $t,g\neq 0.$ (c,d,e) sum of probability densities of the negative $Re(E)$ boundary states for (c)$g=0$, (d) $g>0$, and (e)$g<0.$ }
  \label{fig:chernspectrum}
\end{figure}

To illustrate how we achieve this effect we will use a few examples, the first being a Chern insulator. We choose a simple two-band model for the 2D Chern insulator layer\cite{qi2006topological} having the Bloch Hamiltonian
\begin{equation}
 H_{\text{Ch}}(\mathbf{k}) =  \mathbf{h}(\mathbf{k})\cdot \mathbb{\sigma}
 \end{equation}
where \begin{eqnarray}
\mathbf{h}(\mathbf{k}) = \Bigl( \lambda\text{sin}k_{x},\ \lambda\text{sin}k_{y},\ m + \lambda \sum_{j=x, y} (1-\text{cos}k_{j}) \Bigr),
\end{eqnarray}\noindent and $\sigma$ are the Pauli matrices for spin.
We couple layers in the $z$-direction using the NH terms
\begin{equation}
\begin{aligned}
    H_{\text{layer}} & = \sum_{\alpha=\uparrow,\downarrow}\sum_{k_x,k_y,z}\Bigl[ (t+g)\hat{c}^{\dagger}_{\alpha}(k_x,k_y,z+1)\hat{c}_{\alpha}(k_x,k_y,z)\\
    & + (t-g)\hat{c}^{\dagger}_{\alpha}(k_x,k_y,z)\hat{c}_{\alpha}(k_x,k_y,z+1) \Bigr],   
\end{aligned}
\end{equation} to end up with the full 3D Bloch Hamiltonian:
\begin{equation}\label{eq:H_3D}
 H_{\text{3D}}(\mathbf{k}) =\epsilon(k_z)\mathbb{I}_{2\times 2} + \mathbf{h}(\mathbf{k})\cdot \mathbb{\sigma},
 \end{equation}\noindent where $\epsilon(k_z)=2(t\ \text{cos}k_{z} + ig \text{sin}k_{z})=t_{+}e^{ik_z}+t_{-}e^{-ik_z}.$

Now let us consider the properties of this model. The bulk energy bands in PBC are given by $E_{\pm}(\mathbf{k})=\epsilon(k_z)\pm\vert \mathbf{h}(\mathbf{k})\vert.$  In the Hermitian limit ($g=0$), as we gradually turn on $t$ when $\lambda=1,2<m<4$, the bulk is gapped, and all the topological edge states initially localized on the edges of each decoupled layer will hybridize with each other and spread across the side surfaces. In Fig. \ref{fig:chernspectrum}b we zoom in to show the spectrum of the edge states  (PBC in $z$, open in $x,y$) in the decoupled (blue stars) and weakly-coupled cases (red dots), and show the clearly uniform distribution of the edge states on a $yz$ surface plane in Fig. \ref{fig:chernspectrum}c.  As long as the amplitude of $t$ is not large enough to close the bulk gap, the total number of edge state branches on each surface remains unchanged, i.e., remains equal to the number of $z$-layers, and the system is a 3D weak topological Chern insulator. While this is the regime on which we will focus,  the intermediate coupling regime when $t$ is stronger would result in a Weyl semimetal phase that would also be very interesting to consider in future work.

Now let us consider tuning away from the Hermitian limit.   Because of the NH skin effect we must be careful to study the properties of this system using various combinations of boundary conditions: (i) all directions periodic, (ii) $z$-periodic while one or both of $x,y$ open, (iii) $z$-open while one or both of $x,y$ open. Since the non-Hermiticity in the Bloch Hamiltonian is proportional to the identity matrix, the eigenstates for a periodic system in $z$ are the same for all values of $t,g.$ Hence, for case (i) we simply find a spectrum for the energy bands given in Eq. \ref{eq:H_3D}. For case (ii) we find that the edge state subbands that were broadened along the real energy axis by $t$ will each form complex energy ellipses as shown in Fig. \ref{fig:chernspectrum}a,b. Each ellipse has a non-trivial NH winding number around the region of real energies that were spanned in the $g=0$ limit. Since $z$ is periodic, the system retains its usual topological bulk-boundary correspondence yielding edge states distributed uniformly over the side surfaces, however because of the spectral winding we now know to expect a skin effect for all of the energy states in the spectrum. Indeed, for case (iii), where we expect to see strong effects of the non-Hermiticity,  we find that the spectrum is real, and all the edge states are affected by a skin effect and therefore are exponentially localized on the top (bottom) surface for $g>0$ ($g<0$), as depicted in Fig. \ref{fig:chernspectrum}d,e. This is in sharp contrast to the uniform distribution of the edge states in the Hermitian case shown in Fig. \ref{fig:chernspectrum}c (note the difference of scale in the weight between Fig. \ref{fig:chernspectrum}d,e and \ref{fig:chernspectrum}c).

\begin{figure}[t!]
  \centering
  \includegraphics[width=\columnwidth]{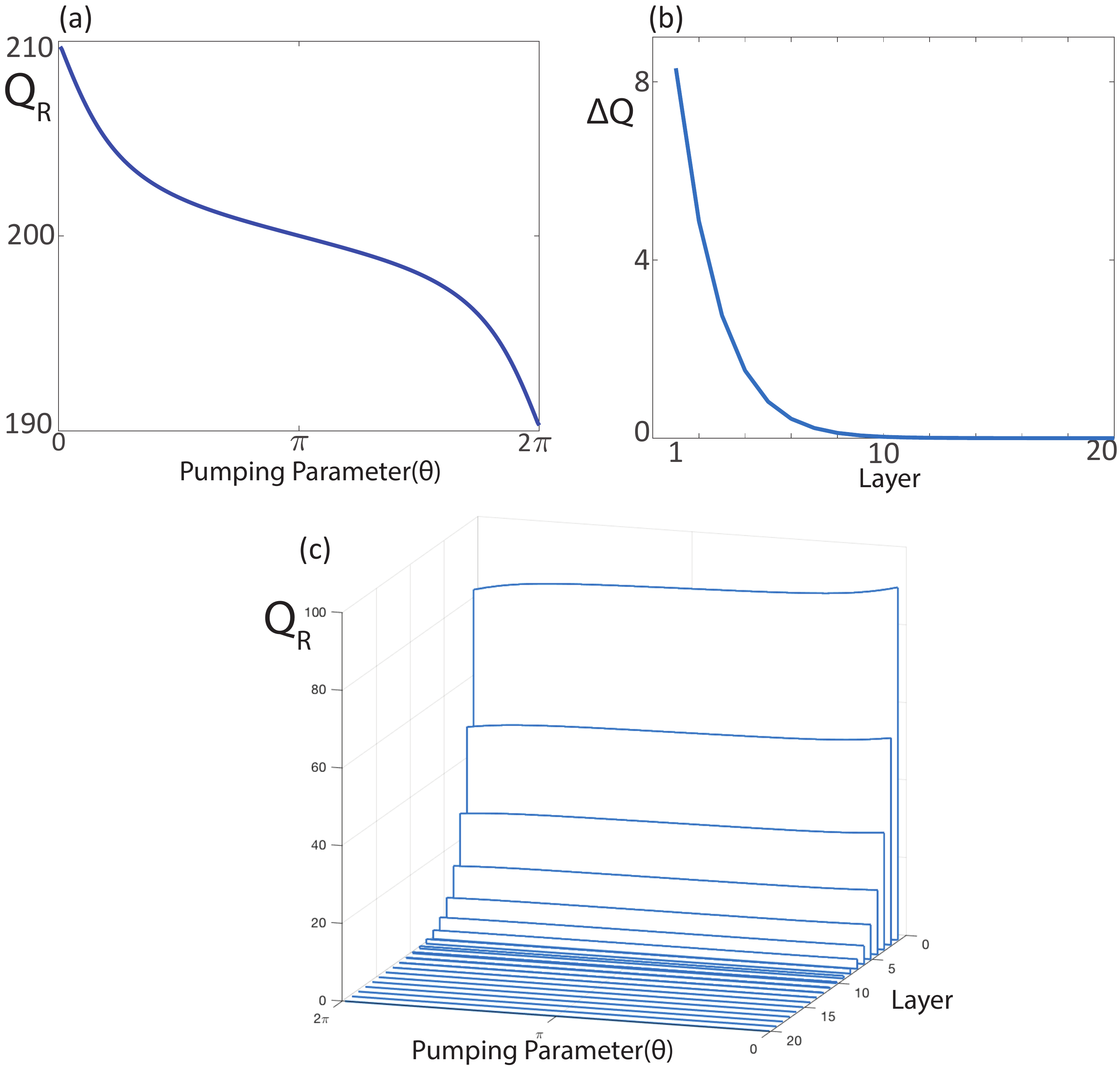}
\caption{(a) Total charge on the right half of the system (positive $x$-direction) as a function of charge pumping parameter $\theta$ (i.e., the flux threaded in the $y$-direction) for twenty Chern insulator layers. (b) Amount of charge pumped on each layer after one cycle. (c) Total chage on the right half of the system resolved vs. layer. Note that the total charge is exponentially localized on one surface, and the variation of charge as a function of $\theta$ is strongest on that surface and decays away.}
\label{fig:pumping}
\end{figure}

In addition to characterizing the spectral features of our model, we will now identify a bulk manifestation of a quantum skin Hall effect via a Laughlin gauge argument\cite{laughlin1981}. To probe the quantum Hall effect we will consider a system having open boundary conditions in $x,z$, but periodic in $y.$ Hence we will thread a flux quantum in the periodic cycle in the $y$-direction. Explicitly, we numerically evaluate the spatially resolved charge density as a function of flux $\Phi=\tfrac{h}{2\pi e}\theta,$ $\theta \in [0,2\pi]$, and then extract the amount of charge pumped in the $x$-direction as a function of flux in the $y$-direction (this is equivalent to applying an electric field in the $y$-direction and measuring the Hall effect through a current in the $x$-direction). For a system with $N_z=20$ Chern insulator layers we find that the total charge transferred to the right side of the system when a single flux quantum is inserted is $|\Delta Q_R|=20$ up to finite-size effects which introduce about a 2$\%$ error (see Fig. \ref{fig:pumping}a). The crucial feature of the quantum skin Hall effect is how this charge transfer is distributed among the layers, as we show in Fig. \ref{fig:pumping}b,c. Indeed, Fig. \ref{fig:pumping}b shows that the amount of charge transferred from one side of the system to the other is exponentially localized near the bottom or top surface depending on the sign of $g.$ Additionally, Fig. \ref{fig:pumping}c shows the total charge on each layer as a function of $\theta.$ In this figure we can see an exponential localization of the total background charge in $z,$ which represents the usual NH skin effect. Crucially, we also see that the flow of charge due to the quantum Hall effect, represented by the variation of the charge on top of the background, is also strongly layer dependent and dies out exponentially quickly as one moves away from the surface harboring the skin effect. The conclusion from the calculations in Figs. \ref{fig:chernspectrum},\ref{fig:pumping}, is that that an extensive number of chiral edge states, and the associated Hall conductance, are localized on the top or bottom surface.


We expect our results to be generalizable to coupled layers of any type of topological system. For example, any topological insulator in 1D/2D that is layered into 2D/3D, and has a Hatano-Nelson coupling between the layers, should exhibit a hybrid bulk-boundary correspondence where topological modes are pushed to the transverse boundary via a NH skin effect. In the Chern insulator case we have seen that chiral edge states pile up on the top or bottom, and since they are $\mathbb{Z}$ classified we expect them to be stable as long as the bulk gap does not close. However, a possible complication may arise if the boundary states are $\mathbb{Z}_2$ classified, e.g., the helical edge modes on a quantum spin Hall(QSH) insulator or the 0D end states of a Su-Schrieffer-Heeger (SSH) type chain\cite{su1980soliton}. We will now argue that such systems have the same stability properties as their corresponding weak topological insulator counterparts, which are known to require a combination of internal symmetries and translation symmetry for protection\cite{fu2007topological}.

\begin{figure}[t!]
  \centering
  \includegraphics[width=\columnwidth]{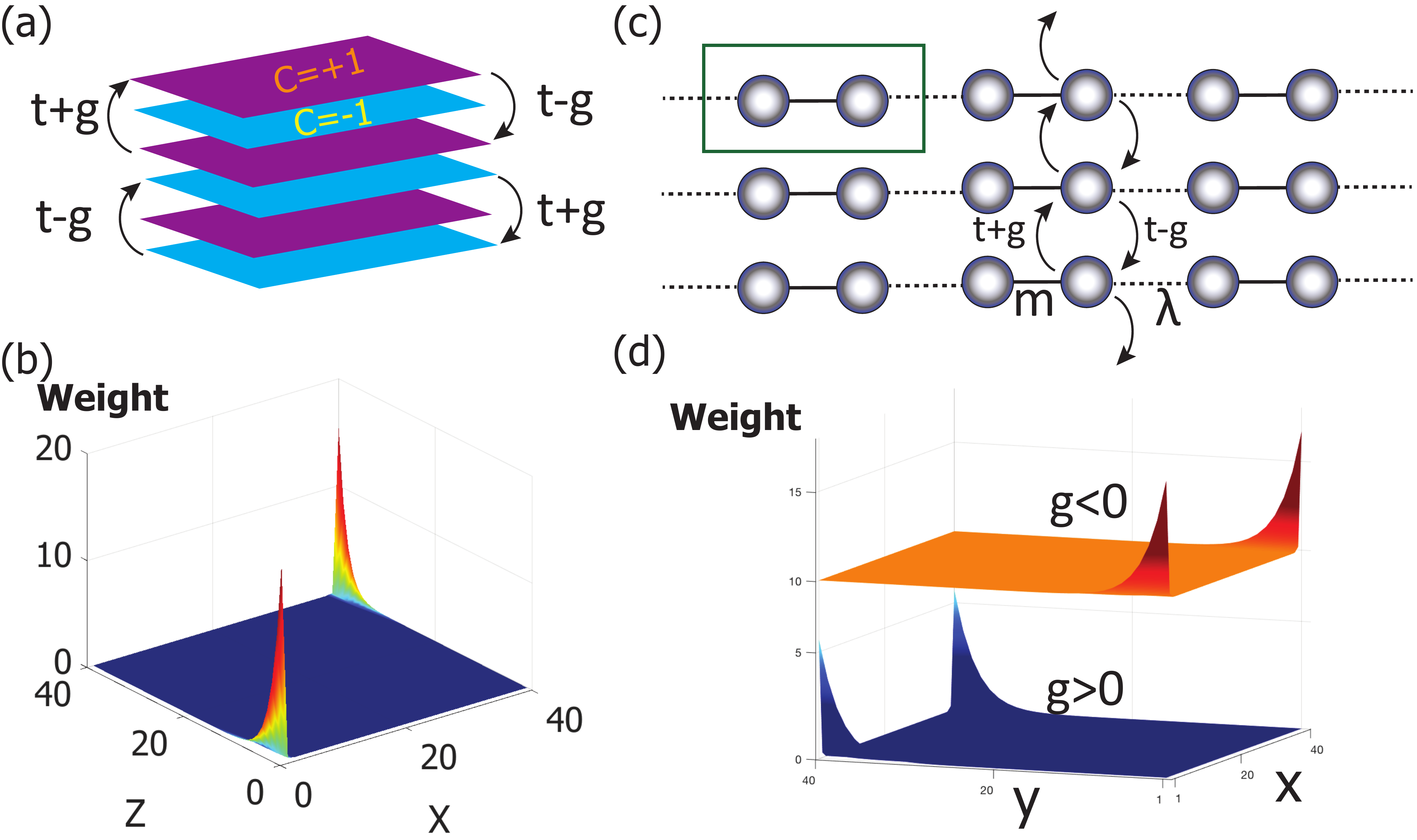}
\caption{(a)Schematic illustration of a $\mathbb{Z}_2$ bi-directional skin effect model built from bi-layers with opposite Chern-number (b) the probability density of the negative energy boundary states at a fixed value of $k_y$ for OBC in $x,z$ and PBC in $y.$ The subspace of boundary states at $-k_y$ would have the corners flipped. (c) Schematic illustration of 2D model built from Su-Schrieffer-Heeger chains parallel to $\hat{x}$ stacked with HN coupling along $\hat{y}.$ (d) Probability density for negative energy boundary modes for OBC in $x,y$ when $g<0$ and $g>0.$ The plot for $g<0$ has been manually shifted upward by a weight of $10$ for easier visibility.}
\label{fig:z2skin}
\end{figure}

To illustrate this, we can stack QSH insulators using a simple four-band model for the 2D TRI topological insulator layer\cite{bernevig2006quantum} having the Bloch Hamiltonian
\begin{equation}
H_{\text{2DTI}}(\mathbf{k}) =  A\text{sin}(k_{x})\Gamma^{1}+ A\text{sin}(k_{y})\Gamma^{2} + M(\mathbf{k})\Gamma^{0},
\end{equation} where $M(\mathbf{k})=M-B[2-\text{cos}(k_{x})-\text{cos}(k_{x})]$,  $\Gamma^{1}=\sigma^{x}\otimes s^{z}$, $\Gamma^{2}=-\sigma^{y}\otimes \mathbb{I}$, $\Gamma^{0}=\sigma^{z}\otimes \mathbb{I}$, and  $\sigma^{i}/s^{i}$ are Pauli matrices acting on orbital/spin space. If we couple $N_z$ layers using the Hatano-Nelson coupling (c.f. Eq. \ref{eq:H_3D}) we will generate an additional term $H=\epsilon(k_z)\mathbb{I}_{4\times 4}+H_{2DTI}(\mathbf{k}).$ Hence the spectrum for this model is effectively two copies of the stack of Chern-insulators we have already studied. For $g=0$ the system is a TRI weak topological insulator with boundary states protected by time-reversal and translation symmetry. For finite $g$ the edge state spectra will form winding loops in complex energy space as in Fig. \ref{fig:chernspectrum}a,b. In this scenario an extensive number of helical modes will manifest on the top/bottom depending on the sign of $g.$ If we break translation symmetry in the $z$-direction, while preserving time-reversal, then we can dimerize the edge states and open a gap using a TRI coupling since they are $\mathbb{Z}_2$ classified. This will give rise to an even-odd effect where $N_z$ being even/odd generically dictates zero/one stable helical edge state that will be pushed to the top or bottom by the NH skin effect. Hence, we find similar stability features to those of a 3D weak topological insulator. We note that we can also induce a bi-directional skin effect (see Fig. \ref{fig:z2skin}a) analogous to Ref. \onlinecite{okuma2020topological}, where we use a modified NH inter-layer coupling to arrive at the Bloch Hamiltonian\begin{equation} H_{bi}(\mathbf{k})=2t\cos(k_z)\mathbb{I}_{4\times 4}+2ig\sin(k_z)\mathbb{I}_{2\times 2}\otimes s^z+H_{2DTI}(\mathbf{k}).\end{equation} This model has a skin effect where an extensive number of edge states of one chirality appear on the top, while the other chirality appear on the bottom, as shown in Fig. \ref{fig:z2skin}b. The modified HN coupling breaks time-reversal and mirror in the $z$-direction, but preserves the product. We find that this effect survives even in the presence of symmetry-preserving terms that couple the opposite spin blocks, though the details of the spin content of the helical skin states will change.

As a final realization of this effect let us consider a system more closely related to experiment. We will focus on a hybrid 2D system where we stack 1D inversion-symmetric SSH chains, and then couple them with the HN coupling (see Fig. \ref{fig:z2skin}c). Both the NH skin effect\cite{xiao2017observation,weimann2017topologically,zhao2018topological,parto2018complex,zhan2017detecting,xiao2020non,helbig2019observation,hofmann2020reciprocal,xiao2020non,zhou2020non,brandenbourger2019non,ghatak2019observation} and the SSH chain\cite{malkova2009,eichelkraut2013,zeuner2015,lee2018topo} have been realized in similar 1D metamaterial systems so their combination should be achievable in the near future.  The Bloch Hamiltonian for this 2D model is
\begin{equation}
    H(\mathbf{k})=\epsilon(k_y)\mathbb{I}_{2\times 2}+(m+\lambda\cos k_x)\sigma^x+\lambda\sin k_x \sigma^y.
\end{equation} We can exhibit the interesting phenomenology of this model by tuning to the strongly dimerized limit $m=0,\lambda=1.$ In periodic boundary conditions this model has the spectrum $E_{\pm}(\mathbf{k})=\epsilon(k_y)\pm 1,$ which, for $t,g\neq 0$ is a pair of complex energy ellipses that are $N_x$-fold degenerate. With open boundaries along $x,$ and with $t=g=0,$ there will be a pair of midgap modes on each SSH chain at zero-energy, and completely localized on the end sites of the chains. Turning on $t,g$ exactly realizes the usual 1D HN chain for the subspace of end-states. Thus, we expect the boundary modes to exhibit a skin effect in the $y$-direction such that for a fully-open, square geometry there will be an extensive number of corner modes on two of the four corners of the square (see Fig. \ref{fig:z2skin}d). This is an unusual feature that would be a smoking-gun spectroscopoic signature for this hybrid topological skin effect phenomenon.

In addition to our newly proposed physical phenomena, we expect that the realization of our models may lead to useful engineering applications. The 1D NH skin effect has already been applied to create a funnel for light\cite{weidemann2020}, and our 2D system of coupled SSH chains could be used as a combination of a signal splitter and a funnel where light impinging upon the system would be split and funneled along the two spatially separated edges. Additionally the extensive number of chiral modes associated to the 3D quantum skin Hall effect system could be used to enhance a topological laser\cite{harari2018,bandres2018}, or as a kind of chiral funnel of light. The hybrid combination of topology transverse to the skin effect opens up a wide design space for new discoveries and applications.

\begin{acknowledgements}
We thank J. Claes and M. R. Hirsbrunner for useful discussions. We thank
the US Office of Naval Research (ONR) Multidisciplinary
University Research Initiative (MURI) grant N00014-20-
1-2325 on Robust Photonic Materials with High-Order Topological Protection for support. We also thank the
US National Science Foundation (NSF) Emerging Frontiers in Research and Innovation (EFRI) grant EFMA1627184 for support.
\end{acknowledgements}

\bibliography{thebibliography.bib}

\end{document}